\documentclass{amsproc}

\theoremstyle{definition}

\theoremstyle{remark}

\numberwithin{equation}{section}

\begin{document}

\title{The D-brane U-scan}

\author{ Eric A.~Bergshoeff}
\address{ Centre for Theoretical Physics,
University of Groningen,  Nijenborgh 4, 9747 AG Groningen, The
Netherlands}
\email{E.A.Bergshoeff@rug.nl}

\author{Fabio Riccioni}
\address{ INFN Sezione di Roma,  Dipartimento di Fisica, Universit\`a di Roma ``La Sapienza'', Piazzale Aldo Moro 2, 00185 Roma, Italy}
\email{Fabio.Riccioni@roma1.infn.it}




\begin{abstract}
We consider the D-branes that occur in IIA/IIB string theory compactified on a torus. We review how a general expression for the Wess-Zumino term of such branes is derived.
We also review the method to determine the D-brane Wess-Zumino term in a U-duality covariant way, and we apply it to derive all the branes obtained by transforming the D-branes under U-duality in any dimension above five. We finally determine all the supersymmetric  branes supporting worldvolume tensor multiplets that occur in these theories in any dimension.
\end{abstract}

\maketitle

\section{Introduction}

D-branes, that are defined as the endpoints of open strings, are a crucial ingredient in the study of non-perturbative string theories.  The fact that string theories possess non perturbative U-duality symmetries implies that D-branes are mapped by these symmetries to other non-perturbative branes. 
In order to study how D-branes transform under U-duality, it is sufficient to consider the low-energy limit, in which this duality can be seen as a continuous symmetry which is the global symmetry of the corresponding supergravity theory. We are interested in theories with maximal supersymmetry, that is IIA/IIB string theory compactified on a torus. For these theories the continuous U-duality symmetries are given in Table \ref{Uduality}.

\begin{table}[t]
\begin{center}
\begin{tabular}{||l|l||}
\hline
dimension $D$&duality group $G$\\
\hline
11&1\\
10A&$\mathbb{R}^+$\\
10B&$\text{SL}(2,\mathbb{R})$\\
9&$\text{GL}(2,\mathbb{R})$\\
8& $\text{SL}(3,\mathbb{R})\times \text{SL}(2,\mathbb{R})$\\
7&$\text{SL}(5,\mathbb{R})$\\
6&SO(5,5)\\
5&$\text{E}_{6(6)}$\\
4&$\text{E}_{7(7)}$\\
3&$\text{E}_{8(8)}$\\
\hline
\end{tabular}
\end{center}
 \caption{\sl   The U-duality groups for all maximal supergravities in
dimensions $3\le D\le 11$. The group is always over the real numbers
and of split real form. In $D=10$ we distinguish between IIA and IIB
supergravity.\label{Uduality}} \end{table}

In general, $p$-branes couple electrically, via a Wess-Zumino (WZ) term, to the $p+1$-form potentials of the low-energy supergravity theory. It turns out that the supersymmetry algebra of maximal supergravity theories closes not only on the propagating gauge fields and their magnetic duals, but also on $D-2$-forms (dual to scalars), $D-1$-forms (dual to constants) and $D$-forms (objects with vanishing field-strength). The allowed U-duality
representations of these potentials have been
classified in \cite{Riccioni:2007au,Bergshoeff:2007qi,deWit:2008ta}. The branes associated to these potentials have co-dimension equal or lower than 2, and as such there are consistency issues that one has to consider in introducing them, most notably the fact that in order to have a finite energy solution one has to introduce orientifolds.\footnote{The U-duality representations of branes with co-dimension higher than two have been extensively studied in the past, see e.g. \cite{Obers:1998rn}.} Nonetheless, one can still determine the number of these branes by only imposing the weaker requirement of local existence, which amounts to requiring the existence of a supersymmetric effective action. Remarkably, this analysis reveals that these ``non-standard'' branes are fewer than the corresponding potentials. 

The counting of non-standard branes was performed in the IIB case in \cite{Bergshoeff:2006gs}. Given the $\text{SL}(2,\mathbb{R})$ covariant forms of the theory \cite{Bergshoeff:2005ac}
\begin{equation}
 A_{2,\alpha}  \ ({\bf 2}) \quad A_{4}   \ ({\bf 1}) \quad A_{6,\alpha}   \ ({\bf 2}) \quad A_{8, \alpha \beta}  \ ({\bf 3}) \quad A_{10,\alpha\beta\gamma}  \ ({\bf 4}) \quad A_{10,\alpha}   \ ({\bf 2}) 
\end{equation}     
one can project on the RR sector, which is associated to D-branes, introducing the charges $q^\alpha$ and $\tilde{q}^\alpha$ such that
 \begin{equation}
B_2 = q^\alpha A_{2,\alpha} \qquad C_2 = \tilde{q}^\alpha A_{2,\alpha}
\end{equation}
are the NS-NS and RR 2-forms. Performing an $\text{SL}(2,\mathbb{R})$ transformation therefore corresponds to swapping $q^\alpha$ and $\tilde{q}^\alpha$. 
 The charge that projects on the RR 8-form is 
  \begin{equation}
  \tilde{q}^\alpha q^\beta q^\gamma q^\delta \epsilon_{\alpha \beta}\quad , \label{charge7braneIIB}
\end{equation}
and denoting with $n(q)$ and $n(\tilde{q})$ the number of $q$'s and $\tilde{q}$'s that occur in a given expression for the charges, we denote the charge in eq. \eqref{charge7braneIIB} with $(n(q), n(\tilde{q})) = (3,1)$. By swapping $q$ and $\tilde{q}$ this charge can only be mapped to the charge (1,3). Consequently, there are two supersymmetric branes inside the triplet of 8-forms $A_{8,\alpha \beta}$.\footnote{The third potential does not correspond to a single supersymmetric brane, but can be associated to  a bound state of two supersymmetric branes \cite{Bergshoeff:2006jj}.} Similarly, the D9-brane corresponds to projecting the quadruplet of 10-forms $A_{10,\alpha\beta \gamma}$ via the charge (4,1), which can only be rotated to the charge (1,4) giving in total two 9-branes inside the quadruplet. This analysis also reveals that the way  the tension scales with the string coupling,
  \begin{equation}
  {\rm Tension} \sim ( g_S )^\alpha \quad , \label{parameteralpha}
\end{equation}
is given by
  \begin{equation}
 \alpha = - n(\tilde{q}) \quad .
\end{equation}

The aim of this contribution is to review the results of \cite{Bergshoeff:2010xc}, where the same techniques where applied to maximal supergravity theories in lower dimensions, leading to a universal gauge-invariant expression for the WZ terms of the D-branes. We will then  apply these methods to determine in any dimension above five the number of non-standard branes that are obtained by performing U-duality rotations on the D-branes. Finally, we will also determine in any dimension the number of 5- branes whose worldvolume degrees of freedom are described by a tensor multiplet. These branes, that we call ``tensor branes'', always belong to different U-duality multiplets with respect to the D-branes.

\section{D-branes in various dimensions}
We wish to repeat the IIB analysis of \cite{Bergshoeff:2006gs} in any dimension $D$ for any maximal supergravity theory. We list the $n$-forms on which the supersymmetry algebra closes as
\begin{equation}
 A_{1 ,M_1} \qquad A_{2, M_2 }  \qquad A_{3, M_3} \qquad ... \qquad A_{D, M_D} \quad , \label{Udualitycovfields}
\end{equation}
where ${M_1 ,M_2 ,...M_D}$ denote the representations of U-duality to which the various forms belong. For all the spacetime-filling $D$-forms and for all forms of rank higher than five  
these representations turn out to be reducible. 

Under gauge transformations, in general each form in \eqref{Udualitycovfields} transforms with respect to the gauge parameters of the lower rank forms according to the gauge algebra of the given theory, whose generalized structure constants are invariant tensors of the U-duality group. For instance, the 2-form $A_{2,M_2}$ transforms as
  \begin{equation}
\delta A_{2,M_2} = d \Lambda_{1,M_2} - f^{M_1 N_1}{}_{M_2}  \Lambda_{0,M_1} F_{2,M_1}
\quad ,
\end{equation}
where $\Lambda_{0,M_1}$ and $F_{2,M_1}$ are the gauge parameter and field strength of the 1-form $A_{1,M_1}$ and $f^{M_1 N_1}{}_{M_2}$ is an invariant tensor. This leads to the gauge-invariant field strength
  \begin{equation}
F_{3,M_2} = d A_{2,M_2} +  f^{M_1 N_1}{}_{M_2} A_{1,M_1} F_{2,M_1}
\quad .
\end{equation}
Similarly, one can define generalized structure constants for higher rank fields, and the closure of the gauge algebra determines the constraints that these invariant tensors have to consistently satisfy  \cite{Riccioni:2009xr,Bergshoeff:2010xc}. 

In any dimension $D$, the U-duality group contains the T-duality subgroup that has the universal expression $\text{SO}(d,d)$, with $d= 10-D$. 
Given that T-duality is a perturbative symmetry of string theory, we expect the D-branes to form representations of the T-duality group. It is therefore useful to decompose the U-duality representations with respect to T-duality according to
 \begin{equation}
 {\rm U-duality} \supset {\text{SO}}(d,d) \times \mathbb{R}^+ \quad .
\end{equation}
In general, for each supersymmetric brane occurring in the theory,
the scaling of its tension with respect to the string coupling, determined by the parameter $\alpha$ in \eqref{parameteralpha}, is related to the ${\mathbb{R}^+}$-weight of the corresponding potential.

Direct inspection in all dimensions \cite{Bergshoeff:2010xc,Bergshoeff:2011zk} reveals a universal structure, at least for the highest values of $\alpha$, which we summarize here:
\begin{itemize}
\item
 Neglecting the scalars, the highest possible value of $\alpha$ is $\alpha=0$, which corresponds to the fundamental fields $B_{1,A}$ and $B_2$ in the vector and singlet representations of T-duality respectively.  These fields are associated to the fundamental particles and the fundamental string.

\item 

The RR fields, that is the fields associated to D-branes, that have $\alpha=-1$, are spinors of T-duality of alternating chirality $C_{2n, \dot{a}}$ and $ C_{2n+1 ,a}$ \cite{Bergshoeff:2010xc}. For standard branes this has been discussed in e.g. \cite{Fukuma:1999jt}.

\item
The solitonic fields, associated to branes with $\alpha=-2$, are antisymmetric tensors of T-duality \cite{Bergshoeff:2011zk}.

\item

The fields associated to branes with $\alpha =-3$ also have a universal structure. It turns out that they belong to tensor-spinor representations of T-duality \cite{wrapping}.

\item

In case of forms belonging to reducible representations, the RR fields (associated to D-branes) always belong to the highest dimensional representation.

\end{itemize}

We are now in the position to generalize the IIB analysis of \cite{Bergshoeff:2006gs} that was reviewed in the introduction.  We introduce two charges $q^{M_1}_A$ and $\tilde{q}^{M_1}_a$ that project the 1-form $A_{1, M_1}$ on the Fundamental and RR 1-forms:
  \begin{equation}
  B_{1, A} = q^{M_1}_A A_{1,M_1} \qquad C_{1,a} = \tilde{q}^{M_1}_a A_{1,M_1} \quad .
 \end{equation}
It follows that the charges that project the U-duality covariant fields \eqref{Udualitycovfields} on the fundamental 2-form $B_2$ and all the RR fields of rank higher than 1 are all given in terms of these charges \cite{Bergshoeff:2010xc}, and using the properties of the invariant tensors of the various U-duality groups it can be shown that these expressions are always unique.  As an example, we write here the expression for the fundamental and RR 2-forms:
  \begin{equation}  
   B_2 = q^{M_1}_A q^{N_1}_B f_{M_1 N_1}{}^{M_2} \eta^{AB} A_{2, M_2}
\qquad  C_{2,\dot{a}} = q^{M_1}_A \tilde{q}^{N_1}_a f_{M_1 N_1}{}^{M_2} \Gamma^A_{\dot{a}}{}^a A_{2, M_2} \quad . 
\end{equation}
Exactly as in the IIB case, the charges that project on the fundamental fields have $n(\tilde{q}) =0$, while the charges that project on the RR fields have $n(\tilde{q})=1$. The general expression for the charge corresponding to the RR fields of rank higher than 2 was derived in \cite{Bergshoeff:2010xc}. 

In order to write down a gauge invariant WZ term for a D-brane, one introduces the world-volume fields $b_{0,A}$ and $b_1$, such that the corresponding gauge invariant field strengths are
  \begin{equation}
{\mathcal F}_{1,A} = d b_{0,A} + B_{1,A} \qquad {\mathcal F}_2 = d b_1 + B_2
\quad .
\end{equation}
It follows that the WZ terms have the universal and  elegant expression
 \begin{equation}
 e^{{\mathcal F}_2}e^{{\mathcal F}_{1,A}\Gamma^A}C  \quad , \label{WZtermgeneral}
\end{equation}
where $\Gamma_A$ are the gamma matrices of the T-duality group $\text{SO}(d,d)$ acting on the various RR fields, which are spinors of T-duality. The above formula thus provides an expression for the WZ term of a D-brane for each component of the T-duality spinor.
Using the properties of these gamma matrices, one can show that the world-volume fields contained in the expression \eqref{WZtermgeneral}, together with the transverse scalars, form the bosonic sector of a world-volume vector multiplet as required \cite{Bergshoeff:2010xc,Bergshoeff:2011zk}.

\section{The D-brane U-scan}

In this section we determine the numbers of different branes obtained by performing U-duality transformations on the D-branes discussed in the previous section. Since for standard branes the result of this analysis is straightforward, because the number of these branes is equal to the dimension of the representation of the corresponding field, we will concentrate on the non-standard branes. The strategy is the same as in the IIB case reviewed in the introduction: given the expression for the charge of the D-brane, which for a RR $n$ form is an expression with $n-1$
$q$'s and one $\tilde{q}$, we perform all possible rotations of the $q$'s and $\tilde{q}$'s. In each expression,  we determine the value of $\alpha$ of the corresponding brane according to 
  \begin{equation}
 \alpha = - n(\tilde{q}) \quad .
\end{equation}
We perform this analysis in any dimension above five. The lower-dimensional cases are more involved and we leave them as an open problem.

\subsection{D=9}
Given that the U-duality symmetry in nine dimensions is $\text{SL}(2,\mathbb{R}) \times \mathbb{R}^+$, the nine-dimensional analysis is similar to the IIB case. The 1-forms are $A_{1,\alpha}$, $A_1$ in the ${\bf 2 \oplus 1}$ of $\text{SL}(2,\mathbb{R})$. We introduce the charges $\tilde{q}^\alpha$, $q^\alpha$ and $q$ such that
  \begin{equation}
  B_1 = q^\alpha A_{1,\alpha} \qquad B_1^\prime = q A_1 \qquad C_1 = \tilde{q}^\alpha A_{1,\alpha} \quad ,
\end{equation}
where $B_1$ and $B^\prime_1$  are the selfdual and anti-selfdual part of a vector of the T-duality group $\text{SO}(1,1)$. 

The charges that project on the RR fields of higher rank are obtained in terms of these charges according to the following rule: going from even to odd rank corresponds to adding a charge $q^\alpha$, and going from odd to even corresponds to adding a charge $q$. For example, the charge of a D1-brane is $\tilde{q}^\alpha q$, while the charge of a D2-brane is $\epsilon_{\alpha\beta} \tilde{q}^\alpha q^\beta q$, which shows  that while the first can be related to a fundamental string of charge $q^\alpha q$, the second is a U-duality singlet. 
Proceeding this way, one can determine all the charges that project onto the RR fields. For the D6-brane one obtains
  \begin{equation}
\tilde{q}^\alpha q^\beta q^\gamma q^\delta q^3 \epsilon_{\alpha\beta} \quad ,
\end{equation}
for the D7-brane one gets
  \begin{equation}
\tilde{q}^\alpha q^\beta q^\gamma q^\delta q^4 \epsilon_{\alpha\beta} 
\end{equation}
and finally for the D8-brane one gets
  \begin{equation}
\tilde{q}^\alpha q^\beta q^\gamma q^\delta q^\epsilon  q^4 \epsilon_{\alpha\beta}  \quad .
\end{equation}
It is instructive to label these charges as $(n(q), n(\tilde{q}))$ in terms of the number of $q^\alpha$ and $\tilde{q}^\alpha$ that occur (the number of $q$ charges is not relevant because it is invariant under U-duality). In particular, the charges of the D6-brane and D7-brane are labelled by (3,1), while the charge of the D8-brane is labelled by (4,1). Performing a U-duality transformation corresponds to swapping $q^\alpha$ and $\tilde{q}^\alpha$, which takes (3,1) to (1,3) and (4,1) to (1,4). This implies that for instance one cannot get a brane with $n(\tilde{q})=2$. The result, giving the full set of non-standard branes in nine dimensions, is summarized in Table \ref{ninedimtable}.

\begin{table}
\begin{center}
\begin{tabular}{|c|c||c|c|c|c|}
\hline \rule[-1mm]{0mm}{6mm} brane & U repr & $\alpha=-1$ & $\alpha=-2$ & $\alpha=-3$ & $ \alpha= -4$ \\
\hline 
\hline \rule[-1mm]{0mm}{6mm} 6-brane & $2 \subset {\bf 3}_3$  & $1$ & $-$ & $1$   & \\
\hline \rule[-1mm]{0mm}{6mm} 7-brane & $2 \subset {\bf 3}_4$  & $1$ & $-$  &  $1$ & \\
\hline \rule[-1mm]{0mm}{6mm} 8-brane & $2 \subset {\bf 4}_4$  &
$1$ & $-$ & $- $ & $ 1 $\\ 
 \hline
\end{tabular}
\end{center}
  \caption{\sl The non-standard branes in nine dimensions. The U-duality group is $\text{SL}(2, \mathbb{R}) \times \mathbb{R}^+$. The subscript indicates 
the $\mathbb{R}^+$-weight. \label{ninedimtable}}
\end{table}

\subsection{D=8}
In eight dimensions the U-duality group is $\text{SL}(3,\mathbb{R}) \times \text{SL}(2,\mathbb{R})$ and the T-duality group is $\text{SL}(2,\mathbb{R}) \times \text{SL}(2,\mathbb{R})$ which is indeed isomorphic to $\text{SO}(2,2)$. 
The 1-form $A_{1,Ma}$ is in the $({\bf \overline{3},2})$ of the U-duality group, and is projected on the fundamental and RR fields through the charges 
 \begin{equation}
q^M_{\dot{a}}
 \qquad \tilde{q}^{M} \quad , 
\end{equation}
where $\dot{a}$ denotes the ${\bf 2}$ of the $\text{SL}(2,\mathbb{R})$ inside $\text{SL}(3,\mathbb{R})$. The fundamental and RR 1-forms are 
  \begin{equation}
B_{1, a\dot{a}} = q^{M}_{\dot{a}} A_{1,Ma} \qquad C_{1,a} = \tilde{q}^M A_{1,Ma} \quad .
\end{equation}
The charges projecting on the fundamental string and the D-string are $\epsilon_{MNP} q^N_{\dot{a}} q^P_{\dot{b}} \epsilon^{\dot{a}\dot{b}}$ and $ \epsilon_{MNP} \tilde{q}^N q^P_{\dot{a}}$ respectively, and one can similarly construct the charges for the higher rank forms using invariant tensors of the U-duality group. Given that there is always one and only one independent expression that gives the right charges, we can neglect the U-duality indices, so that for instance the charges for the fundamental and D-string can be schematically written as
  \begin{equation}
 q_{\dot{a}} q_{\dot{b}} \epsilon^{\dot{a}\dot{b}} \qquad  \tilde{q} q_{\dot{a}}
 \quad .
\end{equation}
Moreover, we can label the charge in terms of the three numbers $(n(q_{\dot{1}}), n(q_{\dot{2
}}), n(\tilde{q}))$, so that for instance the charge of the fundamental string is $(1,1,0)$ while the charge of the D-string is the doublet $(1,0,1)$ and $(0,1,1)$. One can easily see that performing $\text{SL}(3,\mathbb{R})$ transformations these charges are mapped into each other. 

\begin{table}[t]
\begin{center}
\begin{tabular}{|c|c||c|c|c|c|}
\hline \rule[-1mm]{0mm}{6mm} brane & U repr &  $\alpha =-1$ & $\alpha=-2$ & $\alpha=-3$ & $ \alpha= -4$ \\
\hline 
\hline \rule[-1mm]{0mm}{6mm} 5-brane & $6 \subset ({\bf 8},{\bf 1})$  & $2 \subset ({\bf
2},{\bf 1})$ & $2 \subset ({\bf 3},{\bf 1})$ & $2 \subset ({\bf 2},{\bf 1}) $  & \\
\hline \rule[-1mm]{0mm}{6mm} 6-brane & $6 \subset ({\bf 6},{\bf 2})$  & $2 \subset ({\bf
1},{\bf 2})$ & $-$  &  $4 \subset ({\bf 3},{\bf 2})$ & \\
\hline \rule[-1mm]{0mm}{6mm} 7-brane & $6 \subset ({\bf 15},{\bf 1})$  &
$2 \subset ({\bf 2},{\bf 1}) $& $-$ & $2 \subset ({\bf 4},{\bf 1}) $ & $2 \subset ({\bf 3},{\bf 1}) $\\ 
 \hline
\end{tabular}
\end{center}
  \caption{\sl  The non-standard branes  of $D=8$ maximal
supergravity. The U-duality symmetry is $\text{SL}(3,\mathbb{R})
\times \text{SL}(2,\mathbb{R})$ and the T-duality is
$\text{SL}(2,\mathbb{R}) \times \text{SL}(2,\mathbb{R})$.
\label{D=8branescan}}
\end{table}

The charge projecting on the RR 3-form is
  \begin{equation}
 \tilde{q} q_{\dot{a}} q_{\dot{b}}\epsilon^{\dot{a}\dot{b}}
\end{equation}
which we label as $(1,1,1)$
and  is  clearly a singlet. This corresponds to a doublet of D2-branes, because the field possesses an $a$ index with respect to the $\text{SL}(2, \mathbb{R})$ not inside $\text{SL}(3, \mathbb{R})$ which  is left understood in our notation.

Proceeding this way one can determine the charge of all the branes in the theory. We now concentrate on the non-standard branes. The D5-brane charge is 
    \begin{equation}
 \tilde{q} q_{\dot{a}} q_{\dot{b}} q_{\dot{c}} q_{\dot{d}}q_{\dot{e}} \epsilon^{\dot{b}\dot{c}} \epsilon^{\dot{d} \dot{e}}
\end{equation}
which is a doublet of charges $(3,2,1)$ and $(2,3,1)$. This can be rotated to the two solitonic branes $(1,3,2)$ and $(3,1,2)$, as well as to the doublet of $\alpha=-3$ branes with charges $(1,2,3)$ and $(2,1,3)$. 

The D6-brane charge is 
    \begin{equation}
 \tilde{q} q_{\dot{a}} q_{\dot{b}} q_{\dot{c}} q_{\dot{d}}q_{\dot{e}} q_{\dot{f}} \epsilon^{\dot{a}\dot{b}} \epsilon^{\dot{c} \dot{d}} \epsilon^{\dot{e}\dot{f}}
\end{equation}
which we label by $(3,3,1)$. This is a doublet of branes because the 7-form field possesses an extra $a$ index. The fact that there are no 2's in this partition of charges  immediately implies 
that there cannot be any solitonic brane because the charge can never be rotated to something with two $\tilde{q}$'s. One can instead rotate to the two charges $(1,3,3)$ and $(3,1,3)$, corresponding to $\alpha=-3$ branes,  which makes a total of four branes due to the extra $a$ index of the field.

We finally consider the 8-forms. The D7-brane charge is 
      \begin{equation}
 \tilde{q} q_{\dot{a}} q_{\dot{b}} q_{\dot{c}} q_{\dot{d}}q_{\dot{e}} q_{\dot{f}} q_{\dot{g}} \epsilon^{\dot{b}\dot{c}} \epsilon^{\dot{d} \dot{e}} \epsilon^{\dot{f}\dot{g}}
\end{equation}
which corresponds to the doublet of charges $(4,3,1)$ and $(3,4,1)$. Again, this cannot be rotated to a solitonic brane, while there are two branes with $\alpha=-3$, that is $(4,1,3)$ and $(1,4,3)$, and two branes with $\alpha =-4$, that is $(1,3,4)$ and $(3,1,4)$.
The overall result is summarized in Table \ref{D=8branescan}.

\subsection{D=7}

We now consider the branes in seven dimensions, where the U-duality is $\text{SL}(5,\mathbb{R})$ and the T-duality is $\text{SL}(4,\mathbb{R})$, which is isomorphic to $\text{SO}(3,3)$. As in the previous subsection we do not write explictly the U-duality indices, and we project the 1-form $A_{1,MN}$ in the ${\bf \overline{10}}$ of $\text{SL}(5,\mathbb{R})$ 
on the ${\bf \overline{4}}$ and ${\bf 6}$ of $\text{SL}(4,\mathbb{R})$ through the charges
\begin{equation}
\tilde{q}_{a5} \qquad \quad q_{ab} \quad ,
\end{equation}
where the index $5$ has been left explicit to make the embedding of the ${\bf {4}}$ of $\text{SL}(4,\mathbb{R})$ inside the ${\bf 5}$ of  $\text{SL}(5,\mathbb{R})$  more transparent.  

 \begin{table}[t]
\begin{center}
\begin{tabular}{|c|c||c|c|c|c|}
\hline \rule[-1mm]{0mm}{6mm} brane & U repr  & $\alpha=-1$ & $\alpha=-2$ & $\alpha=-3$ & $\alpha=-4$\\
\hline 
\hline \rule[-1mm]{0mm}{6mm} 4-brane & $20 \subset{\bf 24}$   &$4 \subset {\bf \overline{4}}$ & $12 \subset {\bf 15}$ & $4 \subset  {\bf 4} $ & \\
\hline \rule[-1mm]{0mm}{6mm} 5-brane & $20\subset {\bf \overline{40}}$  &$4 \subset {\bf {4}}$ & $ 4 \subset{\bf 10} $& $12 \subset {\bf \overline{20}}$ & \\
\hline \rule[-1mm]{0mm}{6mm} 6-brane & $20 \subset {\bf {70}}$ & $4 \subset {\bf
\overline{4}}$ & $-$ & $12 \subset {\bf {36}}$ &
$4 \subset {\bf 10} $\\
 \hline
\end{tabular}
\end{center}
  \caption{\sl The number of non-standard branes in  $D=7$ maximal
supergravity. The U-duality group is $\text{SL}(5, \mathbb{R})$ and
the T-duality group is $\text{SL}(4, \mathbb{R})$.  \label{D=7branescan}}
\end{table}

The charges of the 2-forms are 
  \begin{equation}
Q_{(1,1)}^d = \tilde{q}_{5a} q_{bc} \epsilon^{5abcd}
\end{equation}
and 
\begin{equation}
Q_{(2,0)}^5 = q_{ab}q_{cd} \epsilon^{abcd5}
\end{equation}
which correspond to the D1-brane and the fundamental string respectively. Here we denote with $Q_{(n,m)}$ the charge of an $n+m$-form with $n(q)=n$ and  $n(\tilde{q}) = m$. Since one can rotate any $d$ index to the index $5$, all the 1-branes are connected by U-duality. We can also derive a general rule to determine how $n(\tilde{q})$ changes when we perform such a rotation. The rule is the following: if we rotate an upstairs index from $a$ to $5$, $n(\tilde{q})$ decreases by one unit and viceversa. If we rotate a downstairs index from $a$ to $5$,  $n(\tilde{q})$ increases by one unit and viceversa.

We now move to the 3-forms. The charge for the D2-brane reads
  \begin{equation}
  Q_{(2,1)}{}_e = \tilde{q}_{5a} q_{bc} q_{de} \epsilon^{5abcd} . \label{sevendimD2}
\end{equation}
This charge can be rotated to $Q_{(1,2)}{}_5$ which is the charge for a solitonic brane, in agreement with our rule. 

We can now write down the expression for the charges of the non-standard D-branes. For the D4-brane one gets
  \begin{equation}
  Q_{(4,1)}{}_i{}^5 = \tilde{q}_{5a} q_{bc} q_{de} q_{fg} q_{hi} \epsilon^{5abcd} \epsilon^{efgh5} \quad .
\end{equation}
This charge can be rotated to the one with $\alpha =-2$, 
$Q_{(3,2)}{}_i{}^j$  with $ i \neq j$ ,
which  corresponds to 12 branes,
and  to the $\alpha=-3$ charge
$Q_{(2,3)}{}_5{}^j$ 
in the ${\bf 4}$ of $\text{SL}(4,\mathbb{R})$.

Next we move to the 5-branes. The D5-branes have charge
  \begin{equation}
  Q_{(5,1)}{}^{5,l5} = \tilde{q}_{5a} q_{bc} q_{de} q_{fg} q_{hi} q_{jk} \epsilon^{5abcd} \epsilon^{efgh5}\epsilon^{ijkl5} \quad ,
\end{equation}
which can be rotated to the $\alpha=-2$ charge
 $ Q_{(4,2)}{}^{l,l5} $
corresponding to 4 branes inside the ${\bf 10}$ of $\text{SL}(4,\mathbb{R})$, and to the $\alpha =-3$ charge
  $
  Q_{(3,3)}{}^{l,lm}$, with $ l \neq m $,
corresponding to 12 branes  inside the ${\bf \overline{20}}$ of $\text{SL}(4)$.

Finally, we consider the 6-branes. The charge of the D6-branes is 
  \begin{equation}
  Q_{(6,1)}{}_m{}^{55}  = \tilde{q}_{5a} q_{bc} q_{de} q_{fg} q_{hi} q_{jk} q_{lm}\epsilon^{5abcd} \epsilon^{efgh5}\epsilon^{ijkl5} 
   \end{equation}
in the ${\bf \overline{4}}$ of $\text{SL}(4,\mathbb{R})$.
This cannot be rotated to a charge with $\alpha=-2$, while one can rotate to the $\alpha =-3$ charge $ Q_{(4,3)}{}_m{}^{nn}$,
with $m \neq n$, 
which gives  12 branes inside the ${\bf 36}$ of $\text{SL}(4,\mathbb{R})$, and  to the $\alpha =-4 $ charge 
$
  Q_{(3,4)}{}_5{}^{nn}$  
which gives  four 6-branes inside the ${\bf 10}$ of $\text{SL}(4,\mathbb{R})$. This concludes the seven-dimensional analysis, which is  summarised in Table \ref{D=7branescan}.

\begin{table}
\begin{center}
\begin{tabular}{|c|c||c|c|c|c|c|}
\hline \rule[-1mm]{0mm}{6mm} brane & U repr  & $\alpha=-1$ &
$\alpha=-2$ & $\alpha=-3$ & $\alpha=-4$ & $\alpha=-5$\\
\hline
\hline \rule[-1mm]{0mm}{6mm}
3-brane & $40\subset {\bf {45}}$  & $8 \subset{\bf 8_{\rm V}}$ & $24\subset {\bf 28} $ & $8 \subset {\bf 8_{\rm V}} $&&\\
\hline \rule[-1mm]{0mm}{6mm}
4-brane & $80\subset {\bf {144}}$  & $ 8 \subset {\bf
8_{\rm S}}$ & $32 \subset {\bf 56_{\rm C}}$ &
$ 32\subset {\bf 56_{\rm S}}
$ & $8 \subset {\bf 8_{\rm C}}$ &\\
\hline \rule[-1mm]{0mm}{6mm}
5-brane & $80 \subset {\bf {320}}$  &  $8 \subset {\bf
{8}_{\rm V}}$ & $8 \subset {\bf 35_{\rm V}}$ & $48 \subset {\bf 160_{\rm
V}}$ & $8 \subset  {\bf 35_{\rm
V}} $ & $8 \subset {\bf 8_{\rm V}}$  \\
\hline
\end{tabular}
\end{center}
  \caption{\sl The  non-standard branes of $D=6$ maximal
supergravity. The U-duality group is $\text{SO}(5,5)$ and
the T-duality group is $\text{SO}(4,4)$.\label{D=6branescan}}
\end{table}

\subsection{D=6}
The six-dimensional case can in principle be treated precisely as the higher dimensional ones, but as we will see the analysis is considerably simplified because the U-duality is the orthogonal group $\text{SO}(5,5)$. Given the 1-form $A_{1 ,  \alpha}$ in the ${\bf 16}$ of $\text{SO}(5,5)$, one introduces the charges $q^\alpha_{\dot{a}}$ and $\tilde{q}^\alpha_a$ such that \footnote{Note that in this six-dimensional case we have denoted the fundamental 1-form with a T-duality spinor index instead of a vector index as in all other cases. This is consistent due to triality of $\text{SO}(4,4)$.} 
  \begin{equation}
 B_{1, \dot{a}} = q^\alpha_{\dot{a}} A_{1,\alpha} \qquad C_{1,a} = \tilde{q}^\alpha_a A_{1,\alpha} \quad .
\end{equation}
Therefore, swapping $q$ and $\tilde{q}$ simply corresponds to swapping the chirality of the T-duality spinors. For every $(m+n)$-form, given the charge denoted by $(n(q),n(\tilde{q}))$,  one can always construct the charge in which  $n(q)$ and $n(\tilde{q})$ are interchanged. This, together with the fact that the number of D-branes, solitonic branes and $\alpha=-3$ branes is known in all dimensions \cite{Bergshoeff:2011zk,wrapping}, automatically determines all the other branes.
For the 4-forms, one has 8 D3-branes, corresponding to the charge $(3,1)$, 24 solitonic branes with charge $(2,2)$, and 8 $\alpha=-3$ branes, corresponding to the charge $(1,3)$. 
The 5-forms give rise to 8 D4-branes (charge $(4,1)$), 32 solitonic branes (charge (3,2)), 32 $\alpha=-3$ branes (charge (2,3)) and 8 $\alpha=-4$ branes (charge (1,4)). The 6-forms in the ${\bf 320}$ give rise to 8 D5-branes (charge (5,1)), 8 solitons (charge (4,2)),  48 $\alpha=-3$ branes (charge (3,3)), 8 $\alpha=-4$ branes (charge (2,4)) and finally 8 $\alpha=-5$ branes (charge (5,1)). The full result is summarized  in Table \ref{D=6branescan}.

\section{Tensor branes}

The analysis we have performed leaves aside another class of branes that occur in string theory, namely the supersymmetric branes whose worldvolume degrees of freedom are described by a tensor multiplet. We call these branes ``tensor branes'', and the fact that tensor multiplets only occur in six dimensions implies that tensor branes must have a six-dimensional worldvolume, and thus only occur in supergravity theories in six dimensions or higher.  

By looking at the way the 6-form potentials decompose under T-duality in various dimensions, one can deduce that in dimensions higher than seven such branes can only be solitonic. The counting of solitonic tensor branes in any dimension was performed in \cite{Bergshoeff:2011zk}, and gives one tensor brane in nine dimensions, two in eight dimensions, four in seven dimensions and eight in six dimensions.
 The solitonic branes reveal the rule to determine the charge of these branes in general. The rule is that one has to take the square of the charge of the 3-form. This rule implies that the number of tensor branes equals in all cases the number of 2-branes, and that the only allowed tensor branes have either $\alpha=-2$ or $\alpha=-4$. 

In seven dimensions,  taking the square of the charge \eqref{sevendimD2} one obtains the charge $Q_{(3,1)}{}_{a} Q_{(3,1)}{}_{a}$, describing 4 branes inside the ${\bf \overline{10}}$ of $\text{SL}(4,\mathbb{R})$. This can be rotated to the charge $Q_{(3,2)}{}_{5} Q_{(3,2)}{}_{5}$, which is a singlet tensor brane with $\alpha=-4$. The six-dimenisonal case works in exactly the same way, giving 8 solitonic tensor branes from the square of the D2-brane charge and other 8 $\alpha=-4$ tensor branes from the square of the solitonic 2-brane charge. The results for all dimensions are summarized in Table \ref{tensorbranes}.

\begin{table}
\begin{center}
\begin{tabular}{|c|c||c|c|c|}
\hline \rule[-1mm]{0mm}{6mm} dimension    & U repr & 
$\alpha=-2$ & $\alpha=-3$ & $\alpha=-4$ \\
\hline
\hline \rule[-1mm]{0mm}{6mm}
$D=6$ & $16 \subset {\bf \overline{126}}$ & $8 \subset  {\bf 35_{\rm
S}}$ & $-$ &  $8 \subset  {\bf 35_{\rm
C}} $ \\
\hline \rule[-1mm]{0mm}{6mm}
$D=7$ & $5 \subset {\bf \overline{15}}$ & $4\subset {\bf {\overline{10}}}$  &  $-$& $1 \subset {\bf 1}$ \\
\hline \rule[-1mm]{0mm}{6mm}
$D=8$ & $2 \subset ({\bf 1,3})$ & $2 \subset {\bf (1,3)} $ &&\\
\hline \rule[-1mm]{0mm}{6mm}
$D=9$ & $1 \subset {\bf 1}$ & $1 $ &&\\
\hline
\end{tabular}
\end{center}
  \caption{\sl The tensor branes in any dimension. These branes are 5-branes, and thus they can only exist in dimension six and above. \label{tensorbranes}}
\end{table}

\section{Conclusions}
We have reviewed the analysis of \cite{Bergshoeff:2010xc}, giving a completely general gauge-invariant expression for the WZ terms of all D-branes in any dimensions and a method to construct a U-duality covariant expression for the RR fields of any maximal supergravity theory. We have applied this method to determine the number of branes that are obtained by performing U-duality transformations on the D-branes. 

It is important to emphasize that the number of solitonic (i.e. $\alpha=-2$) branes and the $\alpha=-3$ branes have been derived recently by the authors using other independent methods \cite{Bergshoeff:2011zk,dualdoubledgeometry,wrapping}, and they coincide in all cases with the ones derived here.

\section*{Acknowledgements}
E.B. would like to thank the organizers of the  String-Math 2011 conference for the opportunity to present this work.

\bibliographystyle{amsalpha}

\begin{thebibliography}{99}

\bibitem{Riccioni:2007au}
  F.~Riccioni and P.~C.~West,
  ``The E(11) origin of all maximal supergravities,''
  JHEP {\bf 0707} (2007) 063
  [arXiv:0705.0752 [hep-th]].


\bibitem{Bergshoeff:2007qi}
  E.~A.~Bergshoeff, I.~De Baetselier and T.~A.~Nutma,
  ``E(11) and the embedding tensor,''
  JHEP {\bf 0709} (2007) 047
  [arXiv:0705.1304 [hep-th]].



\bibitem{deWit:2008ta}
B.~de Wit, H.~Nicolai and H.~Samtleben,
  ``Gauged Supergravities, Tensor Hierarchies, and M-Theory,''
  JHEP {\bf 0802} (2008) 044
  [arXiv:0801.1294 [hep-th]].

\bibitem{Obers:1998rn}
  N.~A.~Obers and B.~Pioline,
  ``U duality and M theory, an algebraic approach,''
  arXiv:hep-th/9812139.




\bibitem{Bergshoeff:2006gs}
  E.~A.~Bergshoeff, M.~de Roo, S.~F.~Kerstan, T.~Ort\'\i n and F.~Riccioni,
  ``SL(2,R)-invariant IIB Brane Actions,''
  JHEP {\bf 0702} (2007) 007
  [arXiv:hep-th/0611036].


\bibitem{Bergshoeff:2005ac}
  E.~A.~Bergshoeff, M.~de Roo, S.~F.~Kerstan and F.~Riccioni,
  ``IIB Supergravity Revisited,''
  JHEP {\bf 0508} (2005) 098
  [arXiv:hep-th/0506013];
E.~A.~Bergshoeff, J.~Hartong, P.~S.~Howe, T.~Ort\'\i n and
F.~Riccioni,
  ``IIA/IIB Supergravity and Ten-forms,''
  JHEP {\bf 1005} (2010) 061
  [arXiv:1004.1348 [hep-th]].


\bibitem{Bergshoeff:2006jj}
  E.~A.~Bergshoeff, J.~Hartong, T.~Ort\'\i n and D.~Roest,
  ``Seven-branes and Supersymmetry,''
  JHEP {\bf 0702} (2007) 003
  [arXiv:hep-th/0612072].





\bibitem{Bergshoeff:2010xc}
  E.~A.~Bergshoeff, F.~Riccioni,
  ``D-Brane Wess-Zumino Terms and U-Duality,''
  JHEP {\bf 1011 } (2010)  139.
  [arXiv:1009.4657 [hep-th]].

\bibitem{Riccioni:2009xr}
  F.~Riccioni, D.~Steele and P.~West,
  ``The E(11) origin of all maximal supergravities: The Hierarchy of
  field-strengths,''
  JHEP {\bf 0909} (2009) 095
  [arXiv:0906.1177 [hep-th]].





\bibitem{Bergshoeff:2011zk}
  E.~A.~Bergshoeff, F.~Riccioni,
  ``String Solitons and T-duality,''
  JHEPA,1105,131.\ 2011 {\bf 1105 } (2011)  131.
  [arXiv:1102.0934 [hep-th]].

\bibitem{Fukuma:1999jt}
  M.~Fukuma, T.~Oota and H.~Tanaka,
  ``Comments on T dualities of Ramond-Ramond potentials on tori,''
  Prog.\ Theor.\ Phys.\  {\bf 103} (2000) 425
  [arXiv:hep-th/9907132].




\bibitem{wrapping}
  E.~A.~Bergshoeff and F.~Riccioni,
  ``Branes and wrapping rules,''
  arXiv:1108.5067 [hep-th].

\bibitem{dualdoubledgeometry}
  E.~A.~Bergshoeff and F.~Riccioni,
  ``Dual doubled geometry,''
  Phys.\ Lett.\  B {\bf 702} (2011) 281
  [arXiv:1106.0212 [hep-th]].




\end{thebibliography}

\end{document}